\documentclass[aps,prl,reprint,showpacs]{revtex4-1}
\usepackage{graphicx}
\usepackage{amssymb}
\usepackage{amsmath}
\usepackage{bm}
\usepackage{verbatim}
\usepackage[bookmarks=false,colorlinks=true,urlcolor=blue,citecolor=blue,linkcolor=blue]{hyperref}
\usepackage[normalem]{ulem}

\begin{document}

\title{Quantum Inverse Freezing and Mirror-Glass Order}

\author{Thomas Iadecola}
\affiliation{Condensed Matter Theory Center and Joint Quantum Institute, Department of Physics, University of Maryland, College Park, Maryland 20742, USA}

\author{Michael Schecter}
\affiliation{Condensed Matter Theory Center and Joint Quantum Institute, Department of Physics, University of Maryland, College Park, Maryland 20742, USA}

\date{\today}

\begin{abstract}
It is well-known that spontaneous symmetry breaking in one spatial dimension is thermodynamically forbidden at finite energy density. Here we show that mirror-symmetric disorder in an interacting quantum system can \emph{invert} this paradigm, yielding spontaneous breaking of mirror symmetry \emph{only} at finite energy density and giving rise to ``mirror-glass" order. The mirror-glass transition, which is driven by a finite density of interacting excitations, is enabled by many-body localization, and appears to occur simultaneously with the localization transition.
This counterintuitive manifestation of localization-protected order can be viewed as a quantum analog of inverse freezing, a phenomenon that occurs, e.g., in certain models of classical spin glasses.
\end{abstract}

\maketitle

The absence of thermal phase transitions for short-ranged interacting one-dimensional (1D) systems follows from the no-go theorems of Mermin-Wagner~\cite{Mermin66} and Hohenberg~\cite{Hohenberg67} for continuous symmetries. For 1D systems with a discrete symmetry, the proliferation of mobile domain walls also precludes symmetry breaking at finite temperatures $T>0$, as occurs, e.g., in the classical Ising chain.  Conventional wisdom based on the eigenstate thermalization hypothesis (ETH)~\cite{Deutsch91,Srednicki94} dictates that these no-go theorems also apply in generic (interacting) isolated quantum systems upon replacing the temperature $T$ by the energy density of an eigenstate: $T\to 0^{\pm}$ corresponds to the ground (ceiling) state at the minimum (maximum) energy density, while $T=\infty$  generally corresponds to highly excited states separated by finite energy densities from both the ground and ceiling states.

Recent work on disordered, interacting quantum systems has shown that some of these no-go theorems may be avoided through the mechanism of many-body localization (MBL), where thermalization fails owing to the existence of extensively many local integrals of motion (LIOMs)~\cite{Pal10,Serbyn13,Swingle13,Huse14,Nandkishore15}. This emergent integrability implies that highly-excited MBL eigenstates generically have low entanglement and behave similarly to gapped ground states of clean interacting systems~\cite{Bauer13}, which \emph{can} exhibit long-range order. Owing to this similarity, it has thus been proposed~\cite{Huse13,Bauer13} to use localization as a means to protect quantum order arising from spontaneous symmetry breaking (SSB) \cite{Huse13,Pekker14,Kjall14,Chandran17,Vasseur16,Friedman17,Prakash17}, or topological transitions \cite{Huse13,Bauer13,Chandran14,Slagle15,Potter15,Parameswaran18}, in highly excited states at a finite energy density above the ground state.

Previous examples of localization-protected order occur for both Abelian \cite{Huse13,Kjall14,Pekker14,Chandran17} and non-Abelian \cite{Vasseur16,Friedman17,Prakash17} discrete symmetries. In the Abelian case, the MBL transition can occur independently of SSB, but SSB at finite energy density is possible only if the system is MBL. In the non-Abelian case, it has been argued that the MBL transition must be accompanied by SSB due to the fundamental incompatibility of MBL and the resonant multiplet structure enforced by the non-Abelian symmetry \cite{Potter16,Protopopov17}.
In all cases studied so far, the ordered phase exhibits symmetry breaking in both the ground state (and/or the ceiling state) and excited states above (below) it. MBL thus effectively promotes the long-range order of the ground (ceiling) state to states in the middle of the many-body spectrum.

\begin{figure}[b!]
\includegraphics[width=\columnwidth]{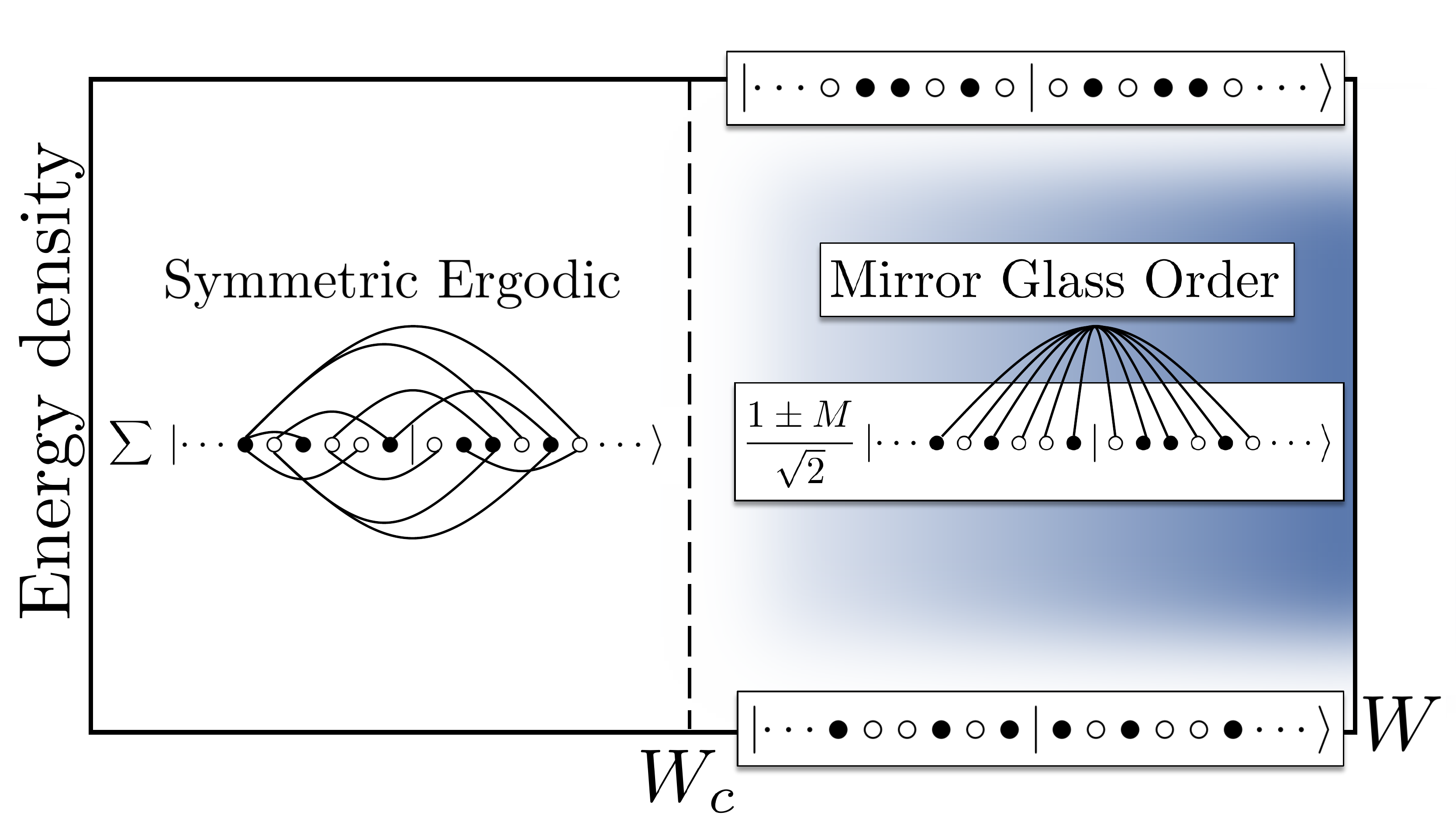}
\caption{Schematic phase diagram of the mirror symmetric random XXZ chain, Eq.~\eqref{eq:H}. At small disorder $W<W_c$ the system is in the symmetric ergodic phase at all energy densities, depicted by a set of highly entangled states. For $W>W_c$ (dashed line) the system becomes many-body localized. This leads to long-range mirror-glass order (depicted in blue) \emph{only} for highly excited mirror-symmetry ``cat" states, which have a \emph{finite} density of excitations above (below) the ground (ceiling) state.}
\label{fig:1}
\end{figure}

In this paper, we introduce a fundamentally different paradigm for SSB in MBL systems in which \emph{neither} the ground state \emph{nor} the ceiling exhibit long-range order, but highly-excited states do.  To exemplify this paradigm, we study a 1D system of interacting particles in a spatially mirror-symmetric disorder potential. The nonlocal nature of this mirror symmetry makes it impossible to construct a complete set of mirror-symmetric LIOMs,
and is therefore manifestly incompatible with MBL. Thus, similar to the case of discrete non-Abelian symmetries, either MBL or the (Abelian) mirror symmetry must break down. We argue that both of these scenarios occur in the same model, the former a quantum-ergodic and mirror-symmetric phase in which the ETH holds, and the latter a localized phase with spontaneously broken mirror symmetry that we dub a ``mirror glass." Unlike SSB in all previously studied MBL models, we find that mirror-symmetry breaking can \emph{only} occur in highly-excited states in the middle of the spectrum. That is, states in the tails of the many-body spectrum remain symmetric through and on either side of the critical point, see Fig.~\ref{fig:1}.

The phenomenon of a symmetric ground state giving way to symmetry-broken states upon increasing energy density has a classical analog in the form of so-called inverse freezing (or inverse melting). In this case, a system that is disordered at zero temperature orders upon \textit{increasing} temperature.  Inverse freezing/melting has been observed in a variety of physical systems~\cite{Schupper05}, and arises in certain models of spin glasses~\cite{Blume66,Capel66,Feeney03,Schupper04}. Unlike previous examples of inverse-freezing phenomena, which occur only at intermediate temperatures, mirror-glass order is inherently quantum (due to MBL) and persists at infinite temperature (i.e., survives averaging over \textit{all} eigenstates, see below). The mirror-glass phase studied here can thus be viewed as the result of a \textit{quantum} inverse-freezing transition.

We study these effects using the random-field XXZ chain, or equivalently, spinless fermions in a 1D random potential described by the Hamiltonian
\begin{align}
H\!=\!&\sum_{i=1}^{L-1} \left[\frac{t}{2}\!\left(c^\dagger_i c_{i+1}\!+\!{\rm H.c.}\right)\!+\!\Delta\! \left(n_i\!-\!\frac{1}{2}\right)\!\left(n_{i+1}\!-\!\frac{1}{2}\right)\right]
\nonumber
\\
&\qquad\!+\! 2\sum_{i=1}^L h_i\! \left(n_i\!-\!\frac{1}{2}\right) ,\label{eq:H}
\end{align}
where $c^\dagger_i/c_i$ are fermion creation/annihilation operators on site $i$, $n_i=c^\dagger_i c_i$ and $L$ is the system length. The Hamiltonian~\eqref{eq:H} respects  mirror symmetry $M$ which transforms site $i\to \bar{i}\equiv L-i+1$, implying $h_i=h_{L-i+1}$ in Eq.~\eqref{eq:H}. A mirror-symmetric disorder potential can be realized using a digital mirror device~\cite{Choi16}; mirror-symmetric \textit{quasiperiodic} potentials can be achieved by phase-locking a pair of incommensurate optical lattices~\cite{Schreiber15,Bordia16}.

We focus on the case of half filling (zero-magnetization sector of the XXZ chain) and consider independent random onsite potentials $h_i$ drawn from a normal distribution with mean zero and standard deviation $W/2$ for $i=1,\dots,L/2$. The potentials for $i=L/2+1,\dots,L$ are determined by the mirror symmetry. We assume the interaction $\Delta$ is smaller than, or on the order of, the hopping $t=1$. In this case, at $W=0$, the ground state is a Luttinger liquid \cite{Giamarchi04} and all excited states are mirror-symmetric. At weak disorder, the system remains in the ergodic phase with mirror-symmetric states at finite energy density obeying ETH. Increasing the disorder strength $W$ leads to MBL. As we show below, in the MBL phase, the ground (ceiling) state and the states at zero energy density above (below) it do not break mirror symmetry spontaneously, while states at finite energy density do.

This behavior is simplest to understand at strong disorder, where one may neglect the particle hopping to leading order.  (Our numerical results below indicate that this reasoning also holds at moderate disorder, which is the case of interest.) In this limit, the eigenstates of the last two terms in Eq.~\eqref{eq:H} consist of product states of the local particle density $n_i$. Their energy is minimized by first populating the sites with the lowest single-particle potential energy $h_i$. Because the potential energy is mirror-symmetric, the lowest-energy eigenstates of $H$ are those with the lowest-energy mirror-related sites \emph{both} populated. Such product states are trivial eigenstates of $M$ with eigenvalue $M=+1$. The same argument applies for states near the top of the spectrum, where only the highest-energy sites are populated. Such states are depicted schematically near the top and bottom edges of the spectrum of Fig.~\ref{fig:1} for $W>W_c$. Here, we represent occupation-basis product states as $|\{n_i\}\rangle=|\circ\bullet\cdots | \cdots \bullet\circ\rangle$, where a closed (open) circle represents an occupied (unoccupied) site and ``$\ |\ $" denotes the mirror axis separating the left and right halves of the system. 

In order to break mirror symmetry spontaneously, one must have a \emph{finite density} of ``excitations," namely pairs of mirror-related sites that are populated by only a single particle, e.g. $|\cdots\circ\circ\bullet\circ\,|\bullet\circ\bullet\bullet\cdots\rangle$. Such product states arise at a finite energy density above (below) the ground (ceiling) states, which are mirror-invariant. They come in degenerate pairs, and are related to their partners by $M$. Of course, eigenstates of $H$ can be labeled by mirror-symmetry eigenvalues $M=\pm1$; this can be achieved by forming the superposition (``Schr\"{o}dinger-cat") states, $|\{n_i\}\rangle_{\pm}=\frac{1\pm M}{\sqrt{2}}|\cdots\bullet\circ\,|\bullet\circ\cdots\rangle$.  These superposition states are only true ``cat states" (i.e., superpositions of \textit{macroscopically different} classical configurations, which lead to SSB) when there is a finite density of mirror-related sites on which $M$ acts nontrivially.  The presence of weak hopping [first term in Eq.~\eqref{eq:H}] leads to an energy splitting between the $|\{n_i\}\rangle_{\pm}$ states that decays exponentially with system size $L$ and vanishes in the thermodynamic limit. This signals the onset of SSB in the highly excited states, which can be chosen as product states with broken mirror symmetry.

The spontaneous breaking of mirror symmetry coincides with the appearance of long-range order in individual eigenstates. Such order can be parameterized by the site polarization $\sigma_i=n_i-n_{\bar{i}}$, which is odd under mirror reflection. In the ergodic phase, the site polarization $\langle\sigma_i\rangle=0$ in every eigenstate. In the symmetry-broken phase, the value of the site polarization is typically finite but \emph{random} in highly excited states. This implies that while $\langle\sigma_i\rangle\neq 0$, the total polarization $\frac{2}{L}\sum_i^{L/2} \langle \sigma_i\rangle=0$ in a generic eigenstate. By analogy with the Edwards-Anderson order parameter for spin glasses~\cite{Edwards75} and its generalizations to the MBL setting~\cite{Kjall14,Vasseur16}, we introduce a mirror-glass order parameter $q$ to capture any long-range correlations present in an eigenstate $|E_n\rangle$,
\begin{equation}\label{eq:q}
q_n=\left(\frac{2}{L}\right)^2\sum_{i,j=1}^{L/2}\langle E_n| \sigma_i\sigma_j|E_n\rangle^2.
\end{equation}
In the ergodic phase, $q_n\to0$ for every eigenstate in the thermodynamic limit, while in the symmetry-broken phase $q_n\to1$ at strong disorder for highly excited states where every pair of symmetry related sites is populated by only a single particle. 

We study the mirror-glass transition using exact diagonalization.  We set the hopping $t=1$ and fix the interaction strength $\Delta = 0.5$ for all numerical calculations, and average our results over at least $650$ disorder realizations and $50$ eigenstates per realization at $L=16$. We also fix the system at half filling, except in the calculations of ground- and ceiling-state properties.  We focus on energies near the center of the many-body spectrum unless otherwise indicated.

\begin{figure}[t!]
\includegraphics[width=\columnwidth]{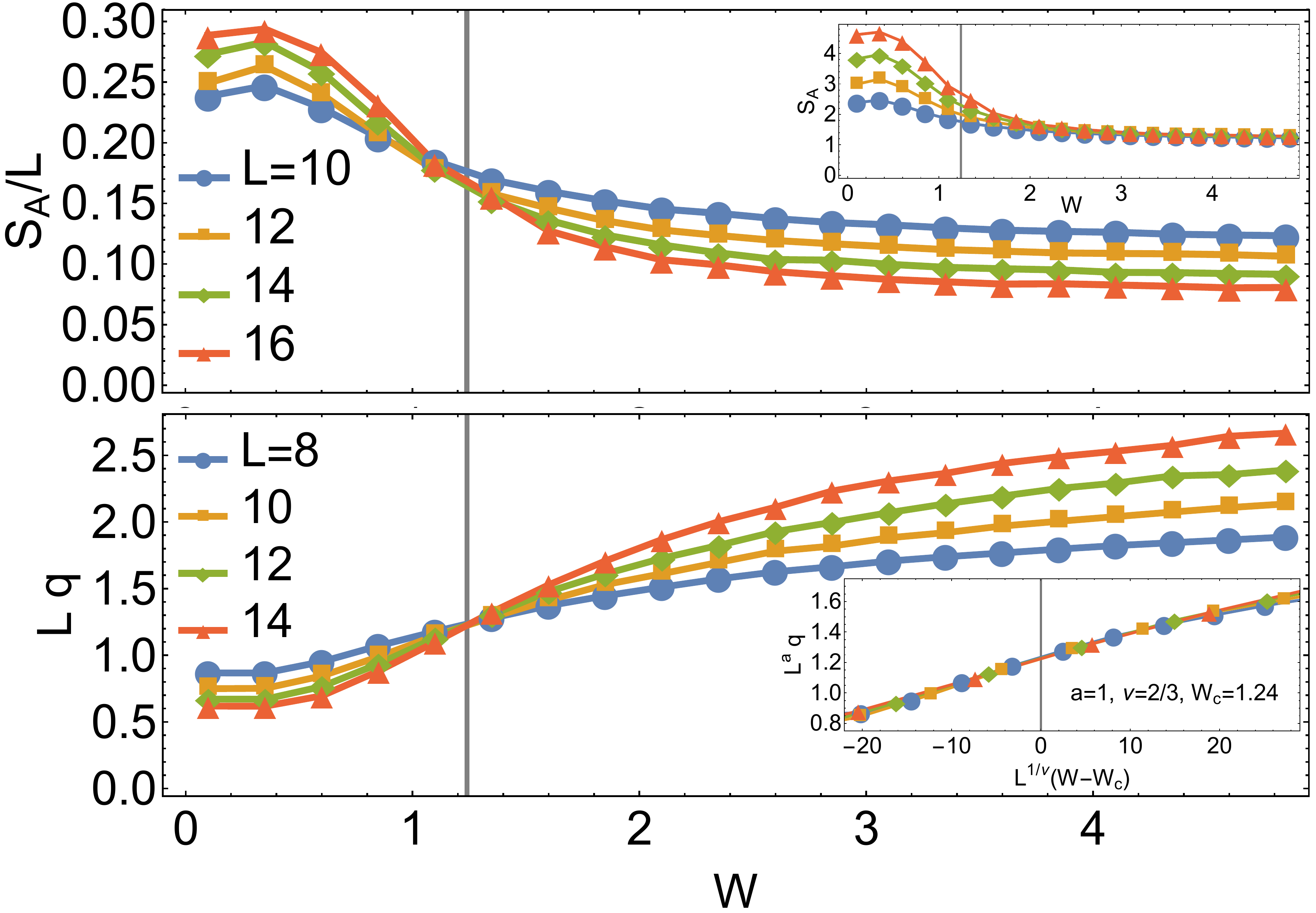}
\caption{Disorder-strength and system-size dependence of the disorder- and eigenstate-averaged bipartite entanglement entropy density $S_A/L$ (upper panel) and mirror-glass order parameter $q$, Eq.~\eqref{eq:q} (lower panel), at $\Delta=0.5$. The average of $q$ is taken over \textit{all} eigenstates, while the average of $S_A$ is taken over the $50$ eigenstates closest to the center of the many-body spectrum. The crossing points at $W_c\approx 1.24$ (grey vertical lines) indicate a transition from the symmetric ergodic phase ($W<W_c$) to the MBL phase ($W>W_c$) with broken mirror symmetry, $q\neq 0$. The inset of the top panel shows $S_A$ (unscaled), to make more apparent the transition between volume law ($W\ll W_c$) and area law ($W\gg W_c$). The inset of the bottom panel shows a scaling collapse of $q$, see Eq.~\eqref{eq:scaling}, near the transition.  }
\label{fig:2}
\end{figure}

In Fig.~\ref{fig:2}, we plot the infinite-temperature average $q=\mathcal{D}^{-1}\sum_n q_n$ as a function of $W$ for various system sizes (here $\mathcal D = L!/(L/2)!^2$ is the Hilbert space dimension at half filling). At small $W$, $q$ tends to zero with increasing $L$, while for $W>W_c$, $q$ develops a finite expectation value. The inset of Fig.~\ref{fig:2} shows a data collapse of $q$ near the critical disorder strength $W_c\approx1.24$ for $\Delta=0.5$. This assumes the scaling form
\begin{equation}\label{eq:scaling}
q=L^{-a}f(L^{1/\nu}(W-W_c)),
\end{equation}
where $a\approx 1$ and $\nu\approx 2/3$. The value of the correlation length exponent violates the Harris-Chayes bound $\nu\geq 2/d$~\cite{Harris74,Chayes86} ($d$ is the dimension of space), but approximately saturates the more general Chandran-Laumann-Oganesyan bound $\nu\geq 2/(d+2a)$~\cite{Chandran15}. Violation of the Harris-Chayes bound could be a consequence of the small systems analyzed here, which would be consistent with other exact-diagonalization results for disordered systems without mirror symmetry~\cite{Kjall14,Luitz15}.

\begin{figure}[t!]
\includegraphics[width=\columnwidth]{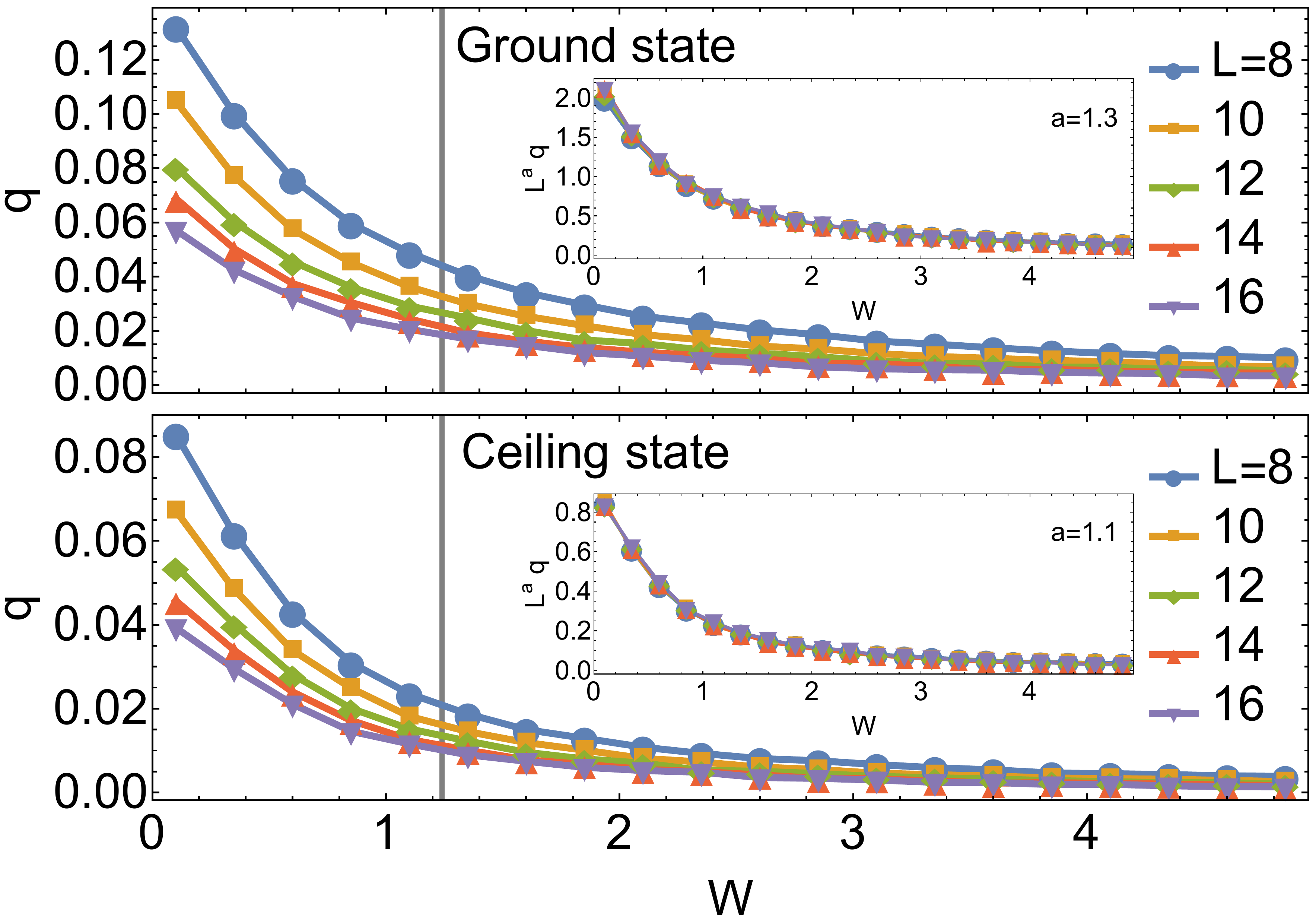}
\caption{Disorder-strength and system size dependence of the disorder averaged mirror-glass order parameter $q$ in the ground state (upper panel) and ceiling state (lower panel). The grey lines again indicate the approximate location of the critical point, $W\approx 1.24$. Insets show a scaling collapse of the order parameter to the form $q=\tilde{q}/L^a$, where $\tilde{q}\sim \mathcal O(1)$ and $a\gtrsim 1$, strongly indicating the absence of mirror-glass order in the tails of the spectrum.}
\label{fig:3}
\end{figure}

To establish MBL, we compute the bipartite entanglement entropy $S_A=-{\rm tr}\rho_A\log\rho_A$, where $\rho_A$ is the reduced density matrix of the half-chain, defined by cutting the bond separating sites $i=L/2$ and $i=L/2+1$.  We average $S_A$ over an energy window near the center of the many-body spectrum, and over disorder realizations. (We consider only the even-inversion sector without loss of generality.)  At strong disorder, we find that $S_A$ tends to a constant of order $\log 2$, consistent with the area-law entanglement expected in an MBL phase~\cite{Bauer13}. At weak disorder, we find that $S_A$ approaches $(L\log 2-1)/2$ as a function of system size, consistent with the ``volume law" characteristic of ergodic quantum systems~\cite{Page93} (see inset of Fig.~\ref{fig:2}, top panel). We plot the disorder- and eigenstate-averaged entanglement entropy density $S_A/L$ as a function of $W$ in Fig.~\ref{fig:2} and find a crossing point near $W\approx 1.24$, consistent with the scaling analysis of the mirror-glass order parameter $q$. These results suggest that the ETH-MBL transition occurs simultaneously with the mirror-glass transition.  

The fact that the two transitions appear to occur at the same critical disorder strength is consistent with the following simple argument based on the assumption that MBL is possible if and only if there exists a complete set of LIOMs~\cite{Serbyn13,Swingle13,Huse14,Imbrie16}.  If such LIOMs exist, then one can enumerate all eigenstates of the MBL system by forming product states with fixed values of the LIOMs on each site.   To construct a \textit{mirror-symmetric} MBL system, we must additionally demand that the LIOMs respect mirror symmetry.  However, because mirror symmetry relates sites $i$ and $\bar{i}$, it is impossible to construct a mirror-symmetric LIOM using only degrees of freedom within, say, a localization length of site $i$.  This means that any \textit{complete} set of LIOMs (and the associated product states corresponding with with MBL eigenstates) necessarily breaks mirror symmetry.

Having demonstrated mirror-glass order in highly excited states, we now illustrate its absence in the spectral tails, which is a hallmark of inverse freezing.  In Fig.~\ref{fig:3}, we plot $q$ for the ground and ceiling states as a function of $W$. One sees that at the critical point $W_c$, and on either side of it, the mirror-glass order tends to zero as $L$ increases. The insets of Fig.~\ref{fig:3} show a collapse of the order parameter when rescaled by the factor $L^a$ with $a\gtrsim 1$. This implies that $q\propto 1/L^a\to0$ as $L\to\infty$, giving strong evidence for the absence of symmetry breaking in the ground and ceiling states. One can verify that a similar analysis applied to any eigenstate of $H$ that has high overlap with a single mirror-symmetric product state would yield the same result.  These states are concentrated in the tails of the many-body spectrum, and form a vanishingly small fraction of the states at finite energy density.

Since Fig.~\ref{fig:2} shows that mirror-glass order occurs at infinite temperature, we conclude that SSB occurs only for generic eigenstates in the middle of the many-body spectrum. The absence of a ground-state quantum critical point distinguishes the mirror-glass phase discussed here from all previous examples of SSB in MBL. In other words, the mirror glass exhibits a novel inverse-freezing effect, where only the highly excited states exhibit long-range mirror-glass order.

Our results strongly suggest the presence of a disorder-driven quantum inverse-freezing transition in the model \eqref{eq:H} with a mirror-symmetric random onsite potential.  Above the critical disorder strength, states in the tails of the many-body spectrum, including the ground and ceiling states, are mirror-symmetric, while generic finite-energy-density states develop mirror-glass order, characterized by the SSB of mirror symmetry.  This unusual pattern of SSB contrasts sharply with all known examples of localization-protected order, where either the ground or ceiling state (or both) exhibit SSB like their counterparts in the bulk of the spectrum.

\begin{figure}[t!]
\includegraphics[width=.6\columnwidth]{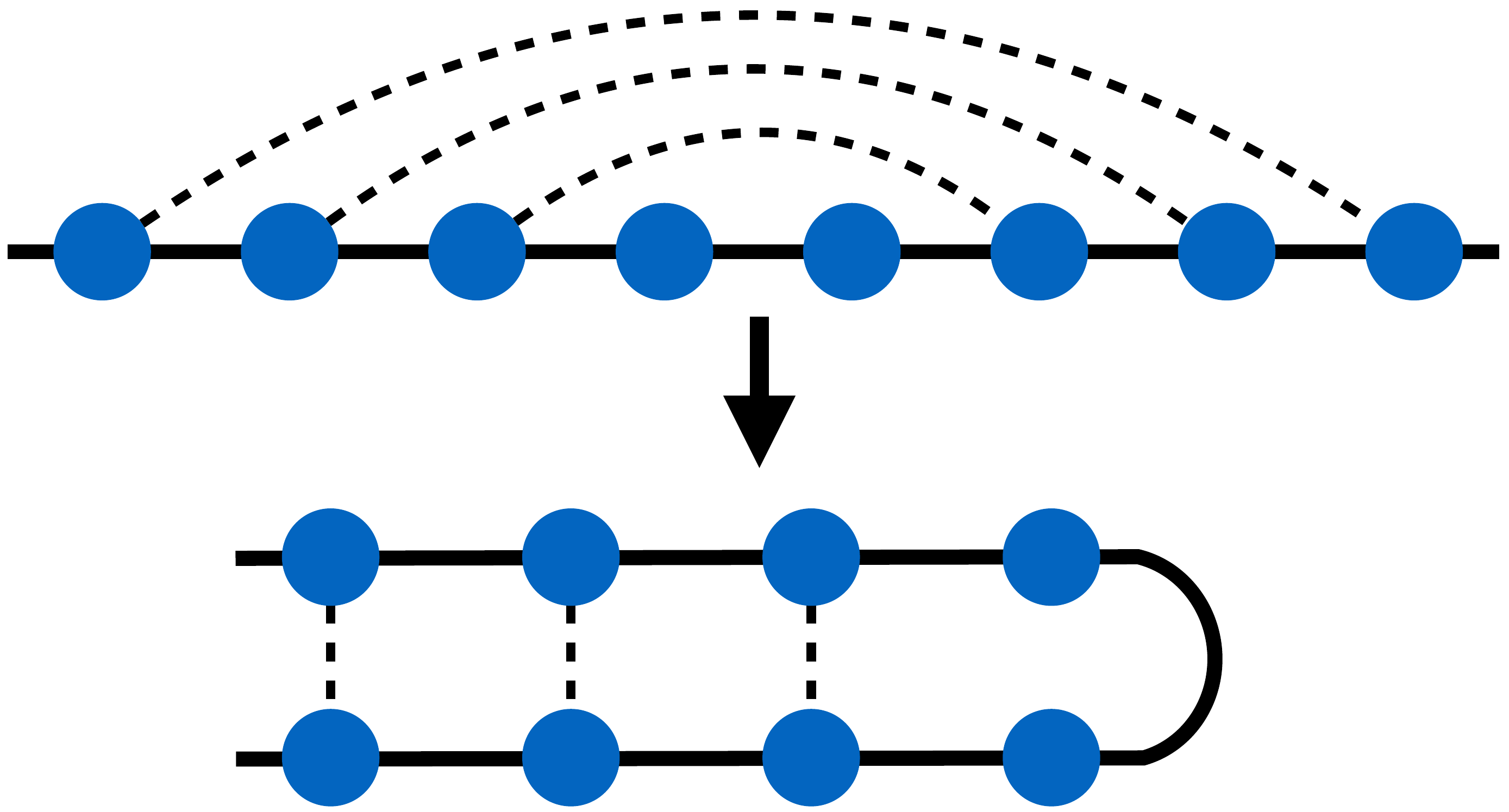}
\caption{Folding a mirror-symmetric chain along the mirror axis into a two-legged ladder.  The solid lines represent nearest-neighbor couplings.  The dashed lines represent nonlocal couplings between mirror-related sites (top) that become local couplings along the ``rungs" of the ladder (bottom).  The locality of such couplings in the folded picture enables the possibility of a symmetry-preserving MBL phase (see main text).
}
\label{fig:4}
\end{figure}

The results reported here open a wide range of interesting problems for future study.  In particular, our results demonstrate that the interplay between strong randomness and spatial symmetries can lead to novel disorder- and interaction-driven quantum phases.  At the same time, the nonlocal nature of the mirror symmetry discussed here is not essential to the physics: for example, one could imagine \textit{folding} the system described by the Hamiltonian~\eqref{eq:H} along the mirror axis, so that mirror symmetry acts in a spatially \textit{local} way (see Fig.~\ref{fig:4}).  The folded system can be viewed as a two-leg ladder (similar ladders have already been realized in cold-atom experiments~\cite{Kaufman16,Bordia16}). One can then investigate the effects of adding local couplings along the ``rungs" of the ladder, which would be highly nonlocal in the unfolded picture.  One important difference relative to the scenario studied in this work is that mirror symmetry in the folded picture is no longer manifestly incompatible with MBL---when the rung couplings are sufficiently strong compared to, e.g., the nearest-neighbor intra-leg interaction $\Delta$, it is possible that the system enters a \textit{symmetric} MBL phase.  However, when the rung couplings are sufficiently weak, the mirror-glass physics studied here likely persists.

It is also worth stressing that, unlike all examples of SSB in MBL systems to date, the inverse-freezing SSB paradigm discussed here does not require random interactions.  Typically, in order for an MBL SSB phase to arise, the random interactions must be \textit{stronger} on average than the random onsite fields~\cite{Kjall14,Pekker14,Vasseur16,Friedman17,Chandran17,Prakash17}, which is likely to be quite difficult to implement experimentally.  In contrast, the mirror-glass example discussed in this paper involves only random onsite potentials, which are easily engineered in quantum simulators.  This suggests that the mirror glass might be a more natural candidate for the experimental observation of SSB in an MBL system.

\acknowledgements{{\it Acknowledgements}.---We thank Sankar Das Sarma for useful discussions and a critical reading of the manuscript.   We acknowledge support from the Laboratory for Physical Sciences and Microsoft. T.I. acknowledges a JQI postdoctoral fellowship.}



\bibliography{refs_iMBL}

\end{document}